# Electromagnetic and Hadron Calorimeters in the MIPP Experiment

T. S. Nigmanov, H. R. Gustafson, M. J. Longo*, H. K. Park, D. Rajaram

*University of Michigan, Ann Arbor, MI 48109, USA*

C. Dukes, L. C. Lu, C. Materniak, K. Nelson, A. Norman

*University of Virginia, Charlottesville, VA 22904, USA*

H. Meyer

*Fermi National Accelerator Laboratory[1], Batavia, IL 60510, USA*

A. Lebedev, S. Seun

*Harvard University, Cambridge, MA 02138, USA*

N. Graf, J. M. Paley

*Indiana University, Bloomington, IN 47405, USA*

G. Aydin, Y. Gunaydin

*University of Iowa, Iowa City, IA 52242, USA*

D. E. Miller

*Purdue University, West Lafayette, IN 47907, USA*

**Abstract.** The purpose of the MIPP experiment is to study the inclusive production of photons, pions, kaons, and nucleons in $\pi$, K, and p interactions on various targets using beams from the Main Injector at Fermilab. The function of the calorimeters is to measure the production of forward-going neutrons and photons. The electromagnetic calorimeter consists of 10 lead plates interspersed with proportional chambers. It was followed by the hadron calorimeter with 64 steel plates interspersed with scintillator. The data presented were collected with a variety of targets and beam momenta from 5 GeV/c to 120 GeV/c. The energy calibration of both calorimeters with electrons, pions, kaons, and protons is discussed. The resolution for electrons was found to be $0.27/\sqrt{E}$, and for hadrons the resolution was $0.554/\sqrt{E}$ with a constant term of 2.6%. The performance of the calorimeters was tested on a neutron sample.

*Corresponding author. Tel. 734 7644445; fax: 734 7639694: e-mail: mlongo@umich.edu

# INTRODUCTION

The MIPP (Main Injector Particle Production) experiment (FNAL E907) [1] took place in the Meson Center beam line at Fermilab. The primary purposes of the experiment were to investigate scaling laws in hadron fragmentation [2]; to obtain hadron production data for the NuMI (Neutrinos at the Main Injector [3]) target to be used for calculating neutrino fluxes; and to obtain inclusive neutron and photon production data to facilitate proton radiography [4]. The data sample collected by MIPP is summarized in Table 1. The electromagnetic and hadron calorimeters allow us to measure the production of forward-going long-lived neutral particles – photons and neutrons – that are not observed in the upstream detectors. The electromagnetic calorimeter was built for the MIPP experiment, while the hadron calorimeter was reused from the HyperCP (E871) experiment [5].

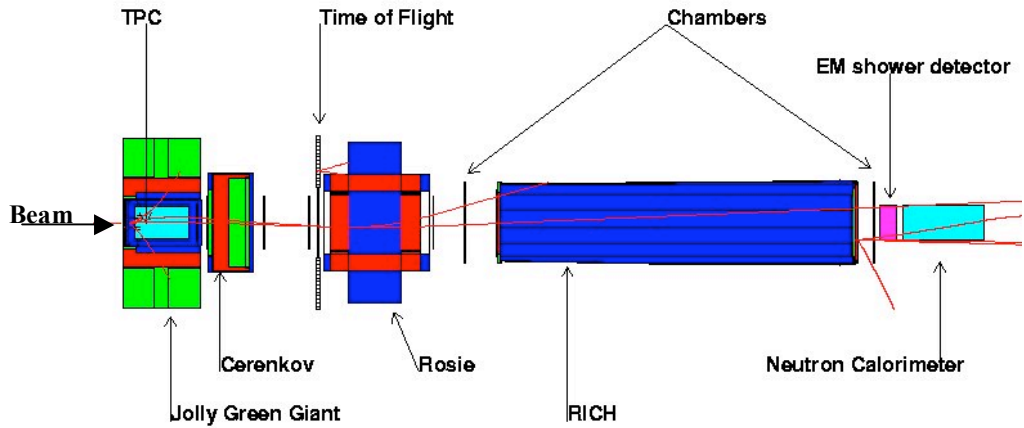

**FIGURE 1**. Experimental layout schematic.

A schematic of the MIPP spectrometers is shown in Fig. 1. The detector consisted of two large aperture magnetic spectrometers. Both magnets had a $p_t$ kick $\cong 0.32$ GeV/c and were operated with opposite polarity so that their deflections approximately canceled. The incident beam entered from the left of the figure and struck targets located ~5 cm upstream of the first magnet. The trajectories and momenta of the secondary charged particles were measured from reconstructed hits in the time projection chamber (TPC), situated inside the first magnet, and hits in the downstream drift chambers and proportional wire chambers. The TPC provided particle identification (PID) in the low energy region (< 1 GeV) by means of ionization (dE/dx); the time-of-flight hodoscope and Cerenkov detector provided PID in the intermediate region (1 – 17

GeV); and the ring-imaging Cerenkov counter (RICH) provided PID for high energy tracks (>17 GeV).

TABLE 1. MIPP data sample.

| Target | Beam Momentum (GeV/c) | Events x $10^6$ |
|---|---|---|
| LH2 | 5, 20, 60 and 85 | 7.08 |
| Beryllium | 35, 60 and 120 | 1.74 |
| Carbon | 20, 60 and 120 | 1.33 |
| NuMI (~0.64 m graphite) | 120 | 1.78 |
| Bismuth | 35, 60 and 120 | 2.83 |
| Uranium | 60 | 1.18 |
| Total | | 15.9 |

# CALORIMETER SPECIFICATIONS

A schematic of the two calorimeters is shown in Fig. 2. The electromagnetic calorimeter (EMCAL) consisted of 10 layers of 5.08 mm thick lead interspersed with planes of gas proportional chambers. The proportional chambers were made from 1.5 m long aluminum extrusions. There were 64 anode wires with 25.4 mm spacing in each plane. The chambers used a gas mix of 76.5% Argon, 8.5% Methane and 15% $CF_4$. The EMCAL active area was 1.6 m wide, 1.5 m high, and 0.3 m in the beam direction. The total thickness was ~10 radiation lengths. The EMCAL pulse height readout system consisted of 640 amplifier channels with multiplexed 12-bit ADCs [6].

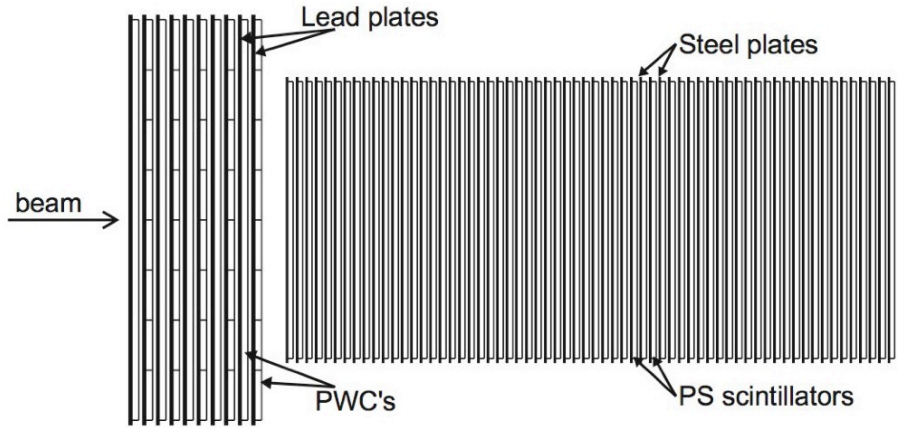

**FIGURE 2**. Schematic of calorimeters (not to scale).

The hadron calorimeter (HCAL) was composed of 64 layers of 24.1 mm iron interspersed with 5 mm thick scintillators as the active medium [5]. The total thickness of the HCAL was 9.6 interaction lengths (88.5 radiation lengths). Its active area was 0.99 m wide, 0.98 m high and 2.4 m in the beam direction. For readout purposes the HCAL was subdivided into four longitudinal and two lateral sections, for a total of 8 cells that were read out with wavelength shifter fibers spaced 30 mm apart. Fibers from each section were bundled into a single 2-inch Hamamatsu R329-02 photomultiplier tube with extended green sensitivity. The pulse heights were flash digitized in custom built CAMAC 14-bit ADC boards with a 75 fC least count.

## ENERGY CALIBRATION

The calibration of the calorimeters was done with incident hadron and electron beams of various momenta. Events were selected with a single charged track in the spectrometer within ~5 cm of the nominal beam position. A match of the EMCAL shower position with the projection of the incoming track was also required. The particles were identified by Cerenkov counters upstream of the target and the RICH downstream of the target.

We assume that the response of each calorimeter is a linear function of the incoming particle's energy so we can write

$$E_i = C_E \sum EMCAL + C_H \sum HCAL \quad (1)$$

where $E_i$ is the particle's energy measured by the upstream magnetic spectrometers, $\Sigma$EMCAL and $\Sigma$HCAL are summed ADC counts representing the EMCAL and HCAL responses for the passage of a given particle, and $C_E$ and $C_H$ are proportionality coefficients to be determined for each beam energy. The dependence of $C_E$ and $C_H$ on beam energy will be discussed in the next section. The $\Sigma$HCAL vs. $\Sigma$EMCAL dependence is illustrated in Fig. 3. In the figure on the left, the large fraction of events near $\Sigma$EMCAL=0 is due to protons that deposit essentially all their energy in the HCAL. As we can see in the plot on the right, in contrast with protons, electrons deposit most of their energy in the EMCAL and little in the HCAL. The few events at small $\Sigma$EMCAL in the electron data are due to hadron contamination in the electron sample. The fit of $\Sigma$HCAL vs. $\Sigma$EMCAL responses is shown in Fig. 4.

The $\Sigma$HCAL vs. $\Sigma$EMCAL dependence appears to be linear for both hadrons and electrons as was assumed in Eq. 1. We can derive the proportionality coefficients from the fit parameters. The calculations of the calibrations coefficients $C_E$ and $C_H$ were done in two ways. In the first method the coefficients were obtained by a least squares fit, and in the second method a maximum likelihood fit was used. The two methods gave consistent results.

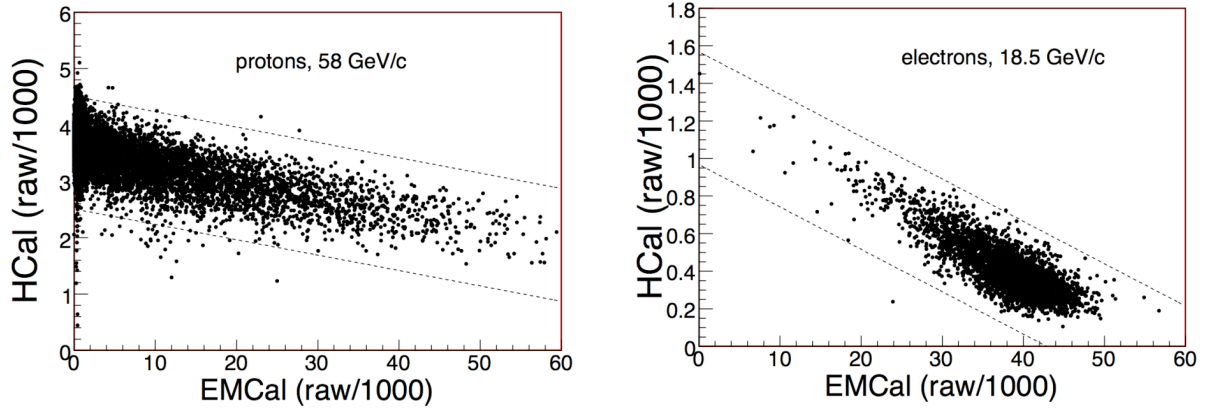

**FIGURE 3**. Scatter plot of ΣHCAL vs. ΣEMCAL responses for the passage of 58 GeV/c protons (on left) and for 18.5 GeV/c electrons (on right). The lines indicate the boundaries for events used in the fitting procedure.

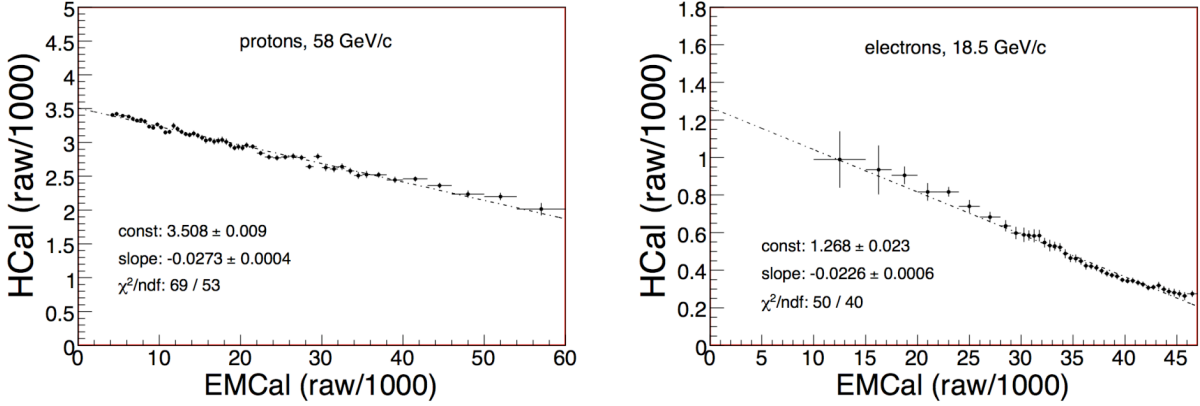

**FIGURE 4**. Averaged ΣHCAL vs. ΣEMCAL responses for 58 GeV/c protons (on left) and for 18.5 GeV/c electrons (on right). The lines represent the results of linear fits.

The final $C_E$ and $C_H$ coefficients are given in Table 2. The errors in the coefficients include systematic uncertainties in the beam momentum. The $C_E$ and $C_H$ coefficients in Table 2 illustrate that the EMCAL and HCAL energy responses at a given energy are quite similar for pions, kaons, protons and antiprotons. The difference in $C_H$ for antiprotons compared to protons is due to the extra 1.88 GeV energy coming from the annihilation of the antiproton. The rest mass energy of the pions and kaons also contributes to small differences in their $C_H$ compared to protons. Since $C_E$ and $C_H$ are correlated through Eq. 1, $C_E$ increases when $C_H$ decreases.

The distributions of $E_{e+h}/p$, the ratio of the energy deposited in the calorimeters to the momentum of the incoming particle, for 58 GeV/c protons and 18.5 GeV/c electrons are shown in Fig. 5. These data validate the calibration procedure and show that the mean values of the $E_{e+h}/p$ ratios are equal to 1. The widths of the distributions reflect the energy resolution of the system.

TABLE 2. The proportionality coefficients $C_E$ and $C_H$ for the EMCAL and HCAL, respectively, calculated for 58 GeV/c $\pi^\pm$, $K^\pm$ and $p^\pm$ and 18.5 GeV/c electrons.

| Particles | p (GeV/c) | $C_E$ (MeV) | $C_H$ (MeV) |
|---|---|---|---|
| $\pi^\pm$ | 58 | 0.46±0.01 | 15.0±0.3 |
| $K^\pm$ | 58 | 0.44±0.01 | 15.3±0.3 |
| $p^+$ | 58 | 0.43±0.01 | 15.5±0.3 |
| $\bar{p}$ | 58 | 0.52±0.02 | 14.9±0.4 |
| $e^-$ | 18.5 | 0.33±0.03 | 14.6±0.3 |

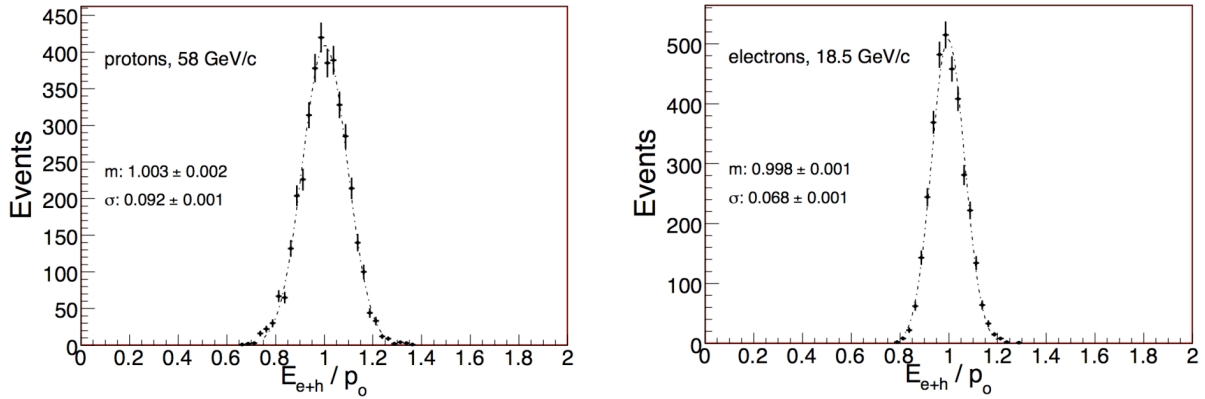

**FIGURE 5**. The distributions of $E_{e+h}/p$ where $E_{e+h}$ is the calorimeter response in energy units, p is the particle's momentum. The left plot shows 58 GeV/c protons data, right plot 18.5 GeV/c electron data. The curves represent the fit results using a Gaussian function.

## ENERGY DEPENDENCE

The energy dependence of the $C_E$ and $C_H$ coefficients was studied using 20, 35, 58, 84 and 120 GeV/c proton data. We were not able to perform the study with electrons since we had electron data only at 18.5 GeV/c. For each proton beam energy we calibrated the EMCAL and HCAL using the procedure described above. The derived $C_E$ and $C_H$ coefficients for the different proton beam energies are given in Table 3. The errors include the uncertainties in the central values of the beam energy. The $C_E$ coefficients in Table 3 decrease with increasing beam energy as a larger fraction of the hadron energy is deposited in the HCAL. The $C_H$ coefficient is less sensitive to the beam energy.

TABLE 3. The proportionality coefficients $C_E$ and $C_H$ for the EMCAL and HCAL, respectively, calculated for 20, 35, 58, 84 and 120 GeV/c protons.

| p (GeV/c) | $C_E$ (MeV) | $C_H$ (MeV) |
|---|---|---|
| 20 | 1.02±0.03 | 16.4±0.4 |
| 35 | 0.78±0.02 | 15.6±0.3 |
| 58 | 0.43±0.01 | 15.5±0.3 |
| 84 | 0.49±0.01 | 14.9±0.3 |
| 120 | 0.52±0.01 | 14.8±0.3 |

# ENERGY RESOLUTION

The energy resolutions ($\sigma/E$) of the calorimeters derived from the widths of the $E_{e+h}/p$ distributions are presented in Table 4. The $\sigma/E$ ratio values have been corrected for contributions from the spectrometer momentum resolution. The combined EMCAL and HCAL energy resolutions were calculated for electrons at 18.5 GeV/c, for $\pi$ and K beams at 58 GeV/c, and for protons at various momenta.

TABLE 4. The combined energy resolution of the calorimeters.

| Particle | p (GeV/c) | $\sigma/E$ (%) |
|---|---|---|
| e | 18.5 | 6.2±0.3 |
| p | 20 | 13.8±1.4 |
| p | 35 | 10.7±0.9 |
| $\pi$ | 58 | 7.6±0.3 |
| K | 58 | 7.6±0.3 |
| p | 58 | 7.6±0.3 |
| p | 84 | 6.7±0.2 |
| p | 120 | 5.9±0.4 |

Recalculating the energy resolution in the form $\sigma/\sqrt{E}$ gives $0.27/\sqrt{E}$ for electrons at 18.5 GeV/c. This value is quite compatible with the energy resolutions of $0.23/\sqrt{E}$ found for electrons using an iron-scintillator calorimeter [7] and $0.33/\sqrt{E}$ using a lead-scintillator calorimeter [8].

The energy resolution of the calorimeters for protons is illustrated in Figure 6. The data points were fitted to $a/\sqrt{E} \oplus b$, where the parameter *a*, which represents statistics-related fluctuations, was found to be 0.554±0.042. The term *b*, which represents detector non-uniformity and calibration uncertainty, was found to be 0.026±0.012. The symbol $\oplus$ indicates addition in quadrature.

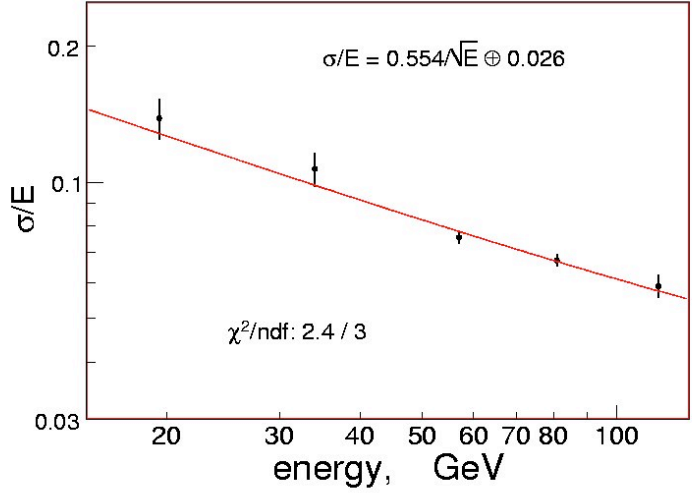

**FIGURE 6**. The energy resolution of the calorimeters as a function of the proton beam energy. The curve represents the fit results using a formula $\sigma/E = a/\sqrt{E} \oplus b$, where E is in GeV.

A comparison of our resolution with that of other calorimeters is shown in Table 5.

**Table 5. Comparison of MIPP energy resolution for protons with other calorimeters.**

| Experiment | Cal Type | Resolution | Resolution @ 20 GeV | Resolution @ 120 GeV |
|---|---|---|---|---|
| MIPP | Pb/Gas+Fe/Scint | $55.4\%/\sqrt{E} \oplus 2.6\%$ | 13.8% | 5.9% |
| D0 [9] | U/LAr | $44\%/\sqrt{E} \oplus 4\%$ | 10.6% | 5.7% |
| FOCUS [10] | Fe/Scint | $85\%/\sqrt{E} + 0.86\%$ | 19.9% | 8.6% |
| HyperCP[1] [5] | Fe/Scint | n/a | n/a | 9% |
| L3 [11] | BGO + U/Gas | $44\%/\sqrt{E} + 7\%$ | 17% | 11% |
| RD-34 [12] | Fe/Scint | $41.3\%/\sqrt{E} + 4.3\%$ | 13.5% | 8.1% |
| WA78 [13] | Fe/Scint | $55\%/\sqrt{E} \oplus 1.7\%$ | 12.4% | 5.3% |
| ZEUS[2] [14] | U/Scint | $43.6\%/\sqrt{E}$ | 9.7% | 4% |
| ZEUS[3] [8] | Pb/Scint | $70\%/\sqrt{E}$ | n/a | 6% |

[1] The HyperCP resolution was measured with 70 GeV protons.

[2] The ZEUS resolution is for all events. The resolution for events fully contained in the calorimeter is 7.8% at 20 GeV and 3.2% at 120 GeV.

[3] ZEUS Forward Neutron Calorimeter. The resolution is for hadrons incident at the center of the tower modules. The resolution is $62\%/\sqrt{E}$ for hadrons incident at the center of the calorimeter.

# ELECTRON-HADRON SEPARATION

We have studied the ability of the calorimeter to separate hadrons and electrons. This was done using beams of electrons at 18.5 GeV/c and hadrons at 20 GeV/c. Figure 7 shows the energy deposited per radiation length for electrons, and per interaction length for hadrons in the EMCAL and HCAL layers. Figure 8 shows of the ratio of the energy in the EMCAL to that in the HCAL. The plot shows that electrons deposit most of their energy in the EMCAL, whereas for hadrons almost all the energy is deposited in the HCAL. We also see that the calorimeters are able to clearly distinguish hadrons from electrons. For example, by requiring that the ratio be >1.0 we are able to separate electrons from hadrons with a hadron rejection ~98.6%.

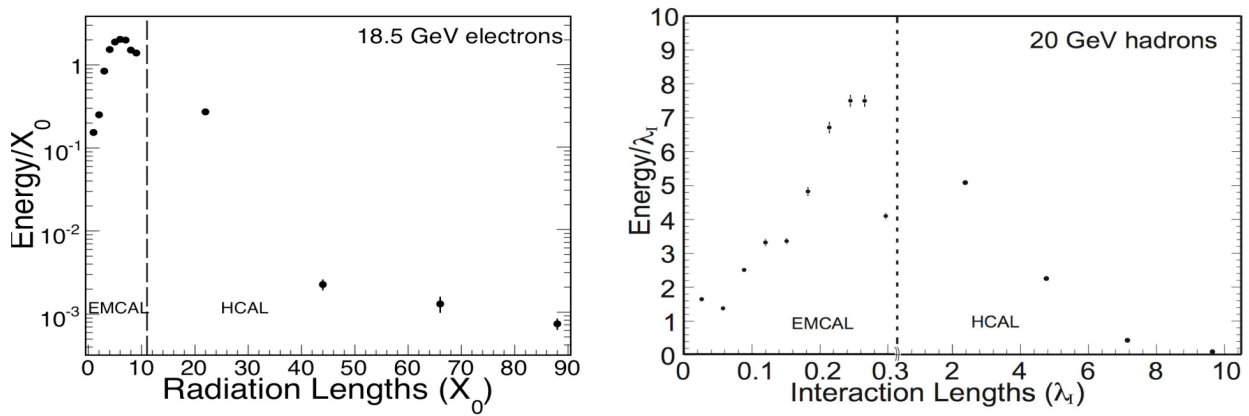

**FIGURE 7.** The energy deposited per radiation length for 18.5 GeV electrons (left) and per interaction length for 20 GeV hadrons (right). The dotted lines represent separation between the EMCAL and the HCAL. Note the change in scale along the x axis between the EMCAL and the HCAL for the figure on the right.

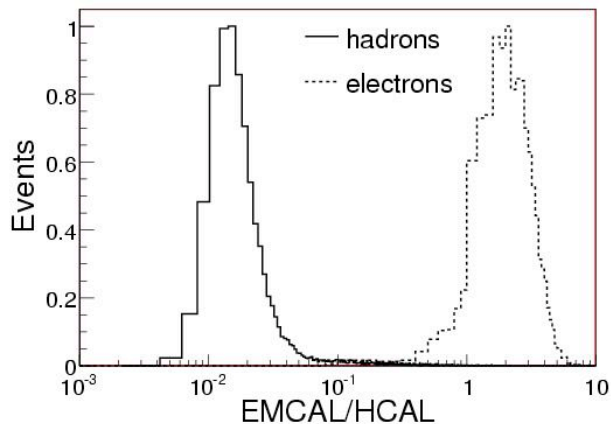

**FIGURE 8.** Ratio of energy in the EMCAL to the energy in the HCAL for 18.5 GeV electrons (dotted line) and 20 GeV hadrons (solid line).

# STUDIES WITH NEUTRONS

The calorimeter performance was also studied using high energy neutrons. The expected signature of neutrons in the calorimeters is: no or a very low energy contribution in the EMCAL (which is about 0.5 interaction lengths) and almost all the energy deposited in the HCAL. A possible source of forward-going high energy neutrons is inclusive charge-exchange production, p+A→n+X. We selected protons using the beam Cerenkov counters and used data from beryllium, carbon, and bismuth targets with thickness 1–2% interaction lengths. For neutron selection we applied a veto on charged high-momentum tracks ($p > 0.3 p_{beam}$). A neutron candidate event with no associated charged track is shown in Fig. 9. The display shows the top

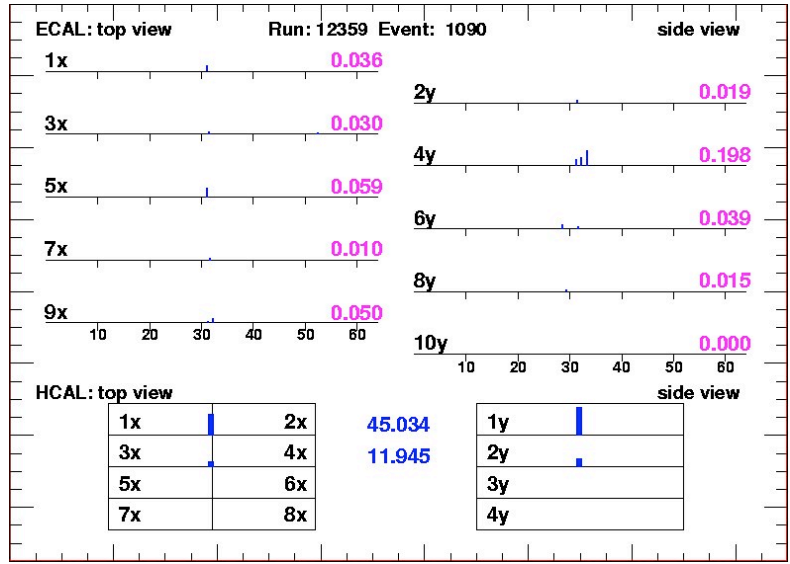

**FIGURE 9**. Calorimeter event display for a charge-exchange candidate event with no associated charged track. The beam momentum is 58 GeV/c, EMCAL energy 0.5 GeV, HCAL energy 57.0 GeV. The beam is incident from the top. Note that the HCAL is not to scale along the beam direction

view (left) and side view (right) of the activity and energy deposited in each layer of the calorimeter. As expected almost all the energy is deposited in the HCAL with minimal (< 1 GeV) deposition in the EMCAL.

The left plot in Fig. 10 shows the energy deposition vs. interaction length for neutrons passing through the EMCAL and HCAL layers. The average total energy deposited in the EMCAL was three orders of magnitude lower than that in the HCAL. The right plot shows the fraction of energy deposited by inclusive neutrons in the calorimeters for 58 GeV and 120 GeV data. The neutron spectra for the two energies appear to be similar. The mean of the distributions is ~60% of the beam energy. The rest of the energy is carried by other particles associated with the neutrons

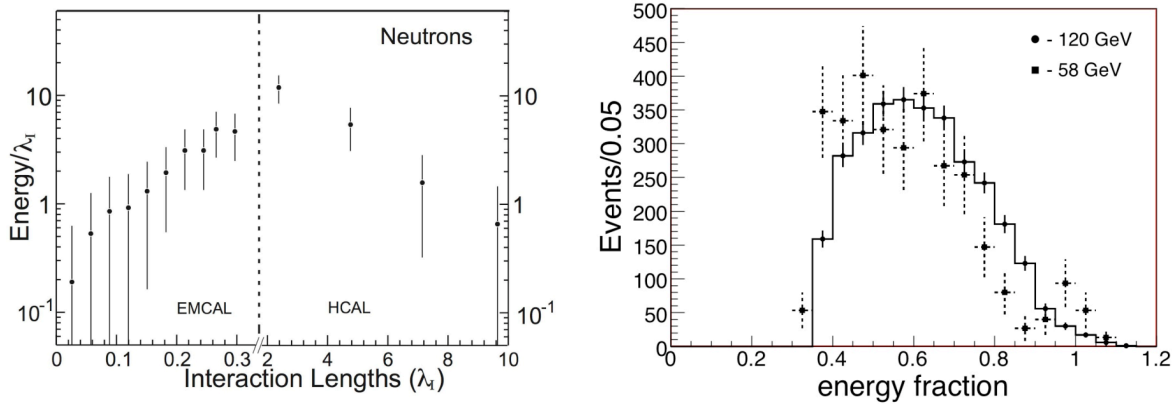

**FIGURE 10**. The left plot shows the energy deposited per interaction length by neutrons passing through the EMCAL and HCAL layers. The right plot shows the energy fraction deposited in the HCAL and EMCAL for neutrons. Note the change in scale along the x axis between the EMCAL and the HCAL for the figure on the left. The data were taken with 58 GeV and 120 GeV proton beams on thin beryllium, carbon, and bismuth targets. Only inclusive neutrons with energy greater than 1/3 of the beam energy are used.

## CONCLUSIONS

The MIPP experiment collected about 16 x $10^6$ events with hadron interactions on various targets with beam momenta from 5 to 120 GeV/c. Charged particles were identified in a wide range of momenta. Use of the electromagnetic and hadron calorimeters made it possible to detect neutral particles and to measure their energies to good accuracy.

The calorimeters were energy calibrated using electrons and hadrons. The EMCAL and HCAL data demonstrate a linear response with energy. The energy calibration was done for electrons at 18.5 GeV/c and for hadrons using 20, 35, 58, 84 and 120 GeV/c beam momenta. For electrons we get a resolution of $0.27/\sqrt{E}$, which is quite close to $0.23/\sqrt{E}$ from [7]. The energy resolution for protons was found to be $0.554/\sqrt{E} \oplus 2.6\%$. Table 5 illustrates that our resolution for protons is comparable to or better than that for other iron or uranium calorimeters except for the ZEUS uranium calorimeter[14]. The calorimeters have the capability of distinguishing electrons from hadrons with a hadron rejection ~98.6%.

The calorimeter was studied on an inclusive neutron sample. It demonstrates the expected response for neutrons: very low energy deposition in the EMCAL and almost all the energy deposited into HCAL.

## ACKNOWLEDGMENTS

The authors express their thanks to colleagues on the MIPP experiment. The efforts of the Fermilab staff are gratefully acknowledged. This research was sponsored by the National Nuclear Security Administration under the Stewardship Science Academic Alliances program through DOE Research Grant DE-FG52-2006NA26182 and the U. S. Department of Energy.

## REFERENCES

1. R. Raja, Nucl. Instr. and Meth. A 553 (2005) 225.

2. R. Raja, Y. Fisyak, in: Proceedings of the DPF92 Meeting, Fermilab.

3. A. G. Abramov, Nucl. Instr. and Meth. A 485 (2002) 209.

4. F. Neri, P. L. Walstrom, Nucl. Instr. and Meth. B 229 (2005) 425;
   John D. Zumbro, Nucl. Instr. and Meth. B 246 (2006) 479.

5. R. A. Burnstein et al., Nucl. Instr. and Meth. A 541 (2005) 516.

6. R. Ball, H. R. Gustafson, M. Longo, T. Roberts, Nucl. Instr. and Meth. 197 (1982) 371.

7. H. Abramowicz et al., Nucl. Instr. and Meth. 180 (1981) 429.

8. S. Bhadra et al., Nucl. Instr. and Meth. A 394 (1997) 121.

9. S. Abachi et al., Nucl. Instr. and Meth. A 324 (1993) 53.

10. V. Arena et al., Nucl. Instr. and Meth. A 434 (1999) 271.

11. C. Chen et al., Nucl. Instr. and Meth. A 272 (1988) 713.

12. F. Ariztizabal et al., Nucl. Instr. and Meth. A 349 (1994) 384.

13. M. De Vincenzi et al., Nucl. Instr. and Meth. A 243 (1986) 348.

14. A. Bernstein et al., Nucl. Instr. and Meth. A 336 (1993) 23.